\documentstyle[preprint,prd,aps]{revtex}
\begin{document}
\thispagestyle{empty}
\draft

\preprint{MSUHEP-50727\, DOE-ER40757-069\, CPP-95-12}

\title{ Production of pseudoscalar Higgs-bosons \\
in $e\,\gamma$ collisions}
\author{Duane A. Dicus}
\address{Center for Particle Physics and Department of 
Physics \\ University of Texas, Austin, Texas 78712}
\author{Wayne W. Repko} 
\address{Department of Physics and Astronomy \\
Michigan State University, East Lansing, Michigan 48824}
\date{\today}
\maketitle

\begin{abstract}
We investigate the production of a pseudoscalar Higgs-boson $A^0$ using the
reaction $e\,\gamma\rightarrow e\,A^0$ at an $e\bar{e}$ collider with center of 
mass energy of 500 GeV. Supersymmetric contributions are included and provide
a substantial enhancement to the cross section for most values of the
symmetry breaking parameters. We find that, despite the penalty incurred in 
converting one of the beams into a source of backscattered photons, the $e\,
\gamma$ process is a promising channel for the detection of the $A^0$.
\end{abstract}
\section{Introduction}
    The Higgs sector in supersymmetric extensions of the standard model contains
charged Higgs-bosons as well as additional neutral Higgs-bosons \cite{hhg}.
Among the latter is a pseudoscalar particle usually denoted $A^0$. In this
paper, we calculate the production cross section for the $A^0$ in the
process $e\,\gamma\rightarrow e\,A^0$. Contributions to this process arise 
from triangle and box diagrams. The triangle contributions consist of 
diagrams in which the $A^0$ and photon are on-shell external particles 
and the remaining particle is a virtual photon or $Z^0$ in the
$t$-channel. Since $t = 0$ is in the physical region, the photon pole
contribution dominates the $Z^0$ pole contribution
in this set of diagrams \cite{scalar}. Moreover, because of the 
off-diagonal structure of the $A^0$ couplings to other bosons, the particles 
in the loop are either quarks, leptons or charginos. Here, we present 
the top quark, bottom quark, tau lepton and the two chargino contributions to
the photon pole amplitude.

The box diagrams have a more
complex particle structure, with leptons, charginos, neutralinos and scalar
leptons in the loops. Like the $Z^0$ pole, these diagrams are non-singular at 
$t = 0$, and should not contribute a sizable correction to the photon pole
terms. They are not included in the present calculation.

\section{The cross section for $A^0$ production}

    The amplitude for the production of an $A^0$ of momentum $k^{\prime}$ and an
$e$ of momentum $p^{\prime}$ in the collision of an $e$ of momentum $p$ and a
$\gamma$ of momentum $k$ and polarization $\varepsilon_{\lambda}(k)$ by the
exchange of a $\gamma$ in the $t$-channel is
\begin{equation} \label{egam}
{\cal M} = \frac{4i\alpha^2}{\sin\theta_W m_W}\bar{u}(p^{\prime})\gamma_{\mu}
u(p)\,\frac{{\cal A}_{\gamma}(t)}{t}\,\varepsilon_{\mu\nu\alpha\beta}
\varepsilon_{\nu}(k)(p - p^{\prime})_{\alpha}k_{\beta}\;,
\end{equation}
where $t = -(p - p^\prime)^2$, and
\begin{eqnarray} \label{amps}
{\cal A}_{\gamma}(t) & = &\;\biggl[-3\left(\frac{2}{3}\right)^2m_t^2\cot\beta\,
C_0(t,m_A^2,m_t^2) - 3\left(-\frac{1}{3}\right)^2m_b^2\tan\beta\,
C_0(t,m_A^2,m_b^2) \\ [6pt] \nonumber
&   &\;\;-(-1)^2m_{\tau}^2\tan\beta\,C_0(t,m_A^2,m_{\tau}^2) + 
2m_Wm_1g_{11}C_0(t,m_A^2,m_1^2) \\ [6 pt]\nonumber
&   &\;\;+\;2m_Wm_2g_{22}C_0(t,m_A^2,m_2^2)\biggr]\,,
\end{eqnarray}
Here, $m_t$ and $m_b$ are the top and bottom quark masses, $m_{\tau}$ is the tau
lepton mass, $m_1$ and $m_2$ are
the chargino masses, $m_W$ is the $W$ mass and $\tan\beta$ is a ratio of 
vacuum expectation values \cite{hhg}. The chargino coupling constants $g_{11}$
and $g_{22}$ depend on the elements of two $2\times 2$ unitary matrices $U$ and
$V$ which diagonalize the chargino mass matrix $X$, where \cite{gh} 
\begin{equation}
X = \left(\!
\begin{array}{cc}
M                    & \sqrt{2}m_W\sin\beta \\
\sqrt{2}m_W\cos\beta & \mu
\end{array}\!\right)\;,
\end{equation}
and are chosen to ensure that $m_1$ and $m_2$ are positive. For illustrative 
purposes,
we assume that the symmetry breaking parameters $M$ and $\mu$ are real and
consider two cases: $M\mu > m_W^2\sin2\beta$ and $M\mu < m_W^2\sin2\beta$. 
The couplings in these cases are
\begin{equation} \label{g1}
g_{11} = \,\frac{m_W}{m_1^2 - m_2^2}(m_2 + m_1\sin2\beta)\,,\quad
g_{22} = \,-\frac{m_W}{m_1^2 - m_2^2}(m_1 + m_2\sin2\beta)\,, 
\end{equation}
for $M\mu > m_W^2\sin2\beta$, and
\begin{equation} \label{g2}
g_{11} = \,\frac{m_W}{m_1^2 - m_2^2}(-m_2 + m_1\sin2\beta)\,,\quad
g_{22} = \,-\frac{m_W}{m_1^2 - m_2^2}(-m_1 + m_2\sin2\beta)\,, 
\end{equation}
for $M\mu < m_W^2\sin2\beta$. Notice that, these couplings are symmetric in 
$m_1$, $m_2$ and,
unlike the $A^0$-top coupling, there is no enhancement factor of $m_{1,2}/m_W$ 
\cite{baer}. We take $m_1 > m_2$. Due to the reality of $M$ and $\mu$, 
$m_1$ and $m_2$ in
Eqs.\,(\ref{g1},\,\ref{g2}) are subject to certain constraints discussed below
\cite{dt}.
The scalar function $C_0(t,m_A^2,m^2)$ is \cite{'thv}
\begin{equation}
C_0(t,m_A^2,m^2) = \frac{1}{i\pi^2}\int d^{\,4}q\frac{1}{\left(q^2 + m^2\right)
\left((q + p - p^\prime)^2 + m^2\right)\left((q + p - p^\prime + k)^2 + m^2
\right)}\;.
\end{equation}
Since one of the external particles is a photon, this function can be
expressed in terms of inverse trignometric or hyperbolic functions 
\cite{hhg,ab-cdr} as
\begin{equation}
C_0(t,m_A^2,m^2)  =\;\frac{1}{(t - m_A^2)}\Bigl(C(\frac{m_A^2}{m^2}) -
C(\frac{t}{m^2})\Bigr)\;,
\end{equation}
where
\begin{eqnarray}
C(\beta) & = &\;\int_0^1\frac{dx}{x}\ln\Bigl(1 - \beta x(1 - x) -
i\varepsilon\Bigr) \\ [6 pt]
         & = &\;\left\{
\begin{array}{lll}
2\left(\sinh^{-1}(\sqrt{-\frac{\mbox{\rule[-3pt]{0pt}{10pt}}\displaystyle\beta}
{\mbox{\rule{0pt}{11pt}}\displaystyle 4}}\,)
\right)^{\!2} &  & \beta\leq 0 \\ [8 pt]
-2\left(\sin^{-1}(\sqrt{\frac{\mbox{\rule[-3pt]{0pt}{10pt}}\displaystyle\beta}
{\mbox{\rule{0pt}{11pt}}\displaystyle 4}}\,)
\right)^{\!2}  &  & 0\leq\beta\leq 4 \\ [8 pt]
2\left(\cosh^{-1}(\sqrt{\frac{\mbox{\rule[-3pt]{0pt}{10pt}}\displaystyle\beta}
{\mbox{\rule{0pt}{11pt}}\displaystyle 4}}\,)
\right)^{\!2} - \frac{\mbox{\rule[-3pt]{0pt}{10pt}}\displaystyle\pi^2}
{\mbox{\rule{0pt}{11pt}}\displaystyle 2} - 2i\pi\cosh
^{-1}(\sqrt{\frac{\mbox{\rule[-3pt]{0pt}{10pt}}\displaystyle\beta}
{\mbox{\rule{0pt}{11pt}}\displaystyle 4}}) &  & \beta\geq 4
\end{array}
\right.\,.
\end{eqnarray}

    The cross section is given by
\begin{equation}
\frac{d\sigma(e\gamma\rightarrow eA_0)}{d(-t)} = \frac{1}{64\pi s^2}
\sum_{\rm spin}|{\cal M}|^2\;,
\end{equation}
and we have
\begin{equation} \label{msq}
\sum_{\rm spin}|{\cal M}|^2 = \frac{\alpha^4}{\sin^2\theta_Wm_W^2}(s^2 + u^2)
\frac{|{\cal A}_{\gamma}(t)|^2}{(-t)}\,,
\end{equation}
where $s = -(p + k)^2$ and $u = - (p^\prime - k)^2$. 
The presence of the $1/t$ in Eq.\,(\ref{msq}) means it is necessary to 
introduce a cutoff in the calculation of the total cross section. One approach
to obtaining a finite cross section is use the 
effective photon or Weizs\"acker-Williams approximation for the exchanged 
photon \cite{eg-gn}. Here, we integrate the exact amplitude and impose an 
angular cutoff. The expression for the total cross section is
\begin{equation} \label{sigma}
\sigma_{e\gamma\rightarrow eA^0}(s) = \frac{\alpha^4}{64\pi\sin^2\theta_W 
m_W^2}\int_{\eta(s - m_A^2)}^{(s - m_A^2)}\frac{dy}{y}\biggl(2 - 
2\frac{(m_A^2 + y)}{s} + \frac{(m_A^2 + y)^2}{s^2}\biggr)|{\cal A}_{\gamma}
(-y)|^2\,,
\end{equation}
where $\eta$ is an angular cutoff. We investigated the effect of varying $\eta =
\sin^2(\theta_{\rm min}/2)$ by comparing the standard model cross section with
and without the $Z^0$ exchange. For $\theta_{\rm min}$ as large as $\pi/6$, the
$Z^0$ contribution is only 3\%-4\% of the total. The result scales approximately
as the logarithm of $\eta$, and we use $\eta = 10^{-5}$ in the figures.

To complete the calculation of the cross
section for the $e\,\gamma$ process, it is necessary to fold the cross section,
Eq.\,(\ref{sigma}), with the distribution $F_{\gamma}(x)$ of backscattered
photons having momentum fraction $x$ \cite{gam} to obtain
\begin{equation}
\sigma_T = \frac{1}{s}\int_{m_A^2}^{0.83s}d\hat{s}F_{\gamma}(\frac{\hat{s}}{s})
\sigma_{e\gamma\rightarrow eA^0}(\hat{s})\;,
\end{equation}
with $\hat{s} = xs$. Here, we have taken the usual upper limit on the allowed
$x$ value, $x =0.83$. 

    This cross section is plotted in Fig.\,(1) for $M\mu > m_W^2\sin2\beta$ and 
in Fig.\,(2) for $M\mu < m_W^2\sin2\beta$. 
The dashed line in each panel is the contribution
from the top and bottom quarks and the tau lepton. For large $\tan\beta$, the
tau contribution is important. This is illustrated in the $\tan\beta = 20$ panel
of Fig.\,(1), where the dot-dashed line is the contribution from the top and
bottom quarks. In Fig.\,(1), the solid lines are $m_1 = 250$ GeV and $m_2$ the
largest value consistent with the constraint $(m_1 - m_2)\geq m_W\sqrt{2(1 +
\sin2\beta)}$, which is needed to ensure that $M$ and $\mu$ are real. Similarly,
in Fig.\,(2), the solid lines correspond to $m_1 = 250$ GeV and $m_2$ the
largest value consistent with $(m_1 - m_2)\geq m_W\sqrt{2(1 - \sin2\beta)}$.
Unlike the $M\mu > m_W^2\sin2\beta$ case, when $M\mu < m_W^2\sin2\beta$ it is 
possible for $m_1$ and
$m_2$ to be equal for $\tan\beta = 1$ provided the $m_1, m_2 \geq m_W$. These
values of $m_1$ and $m_2$ are within the range of chargino masses found in
studies of minimal supersymmetric models \cite{kkrw}.
In most cases, the inclusion of the
chargino contribution leads to a significant increase in the cross section,
especially for the larger values of $\tan\beta$. 

    To assess the observability of this process, we assume that the dominant
$A_0$ decay is $A_0\rightarrow b\bar{b}$. For $m_{A} < 2m_t$, this ignores
some contribution from chargino pair decay, but this is relatively small since
in all but one of the examples we consider the lowest chargino mass exceeds
$\sim 120$ GeV. Even above the top threshold, $b\bar{b}$ decay dominates when
$\tan\beta\sim 20$ \cite{bcps}\,. The dotted lines in Figures (1) and (2) are 
the cross section for the direct production of 
a background $b\bar{b}$ of invariant mass $m_{A}$ in $e \gamma$ collisions
subject to an angular cut on the $b$ and $\bar{b}$ direction relative to the 
that of the incident photon in the $e \gamma$ center of mass.
We find that an angular cut of $|\cos\theta| < .98$ on both the $b$ and the
$\bar{b}$ reduces the background $b\bar{b}$ cross section by about a factor of
10 while leaving the $b\bar{b}$ signal from $A^0$ decay essentially
unchanged. More restrictive cuts on the $b$ and $\bar{b}$ angles can further
suppress the background, but at the expense of a significant decrease in the
signal \cite{Z}. The cut shown appears to be optimal.

\section{Discussion}

    We would like to point out that the $e\,\gamma$ cross sections calculated
here are very likely to be much larger than those of the related process $e\,
\bar{e}\rightarrow \gamma A^0$ at 500 GeV. We have checked this for the
production of the standard model Higgs-boson using the complete (standard model)
calculation of $e\,\bar{e}\rightarrow \gamma H^0$ \cite{ab-cdr} and the photon
pole contribution to $e\,\gamma\rightarrow e\,H^0$. 
At an $e\,\bar{e}$ center of mass energy of 500 GeV, we find the cross section 
$\sigma(e\,\bar{e}\rightarrow \gamma H^0)$ for the production of a 200 GeV 
$H^0$ is 0.08 fb, whereas $\sigma(e\,\gamma\rightarrow e\,H^0) = 5.9$ fb for the
same Higgs-boson mass. 

    This enhancement is implicit in a previous calculation of scalar Higgs-boson
production \cite{eg-gn}. In Ref.\,\cite{eg-gn}, the Weizs\"acker-Williams
approximation is used for the $t$ channel photon together with the on-shell 
$H\rightarrow \gamma\gamma$ amplitude. This is essentially equivalent to setting
$y = 0$ in the parentheses of Eq.\,(\ref{sigma}) and using $m_e^2$ as the cutoff
in the remaining integral \cite{eq1}. Our comparison of the approximate results
of Ref.\,\cite{eg-gn} with an exact calculation suggests that the
Weizs\"acker-Williams approach tends to overestimate the cross section. 
Apart from minor variations depending on how the
calculation is performed, it is nevertheless true that the $t$ channel cross
section is substantially larger than its $s$ channel counterpart.

    The production of $A^0$ has also been investigated in $\gamma\gamma$
collisions \cite{gh1}\,. In this case, the background arises from the process
$\gamma\gamma\rightarrow b\bar{b}$ and it is effectively suppressed by imposing
an angular cut. With our choice of chargino masses, a comparison of the
$\tan\beta = 20$ cross sections in $e \gamma$ production and $\gamma\gamma$
production \cite{gh1} reveals a larger signal in the $e\gamma$ mode. Both
channels are likely to be important in searches for the $A^0$.

    To the extent that the photon pole contribution can be isolated, this method
of searching for the $A^0$ has the advantage that the contributions from
supersymmetry are significant and limited to one type supersymmetric particle.
Should one observe a cross section larger than any standard model prediction,
the case for the presence of chargino contributions is rather strong.

\acknowledgments

We would like to acknowledge conversations with C.-P. Yuan and X. Tata. This
research was supported in part by the National Science Foundation under grant
PHY-93-07980 and by the United States Department of Energy under contract
DE-FG013-93ER40757.

\newpage
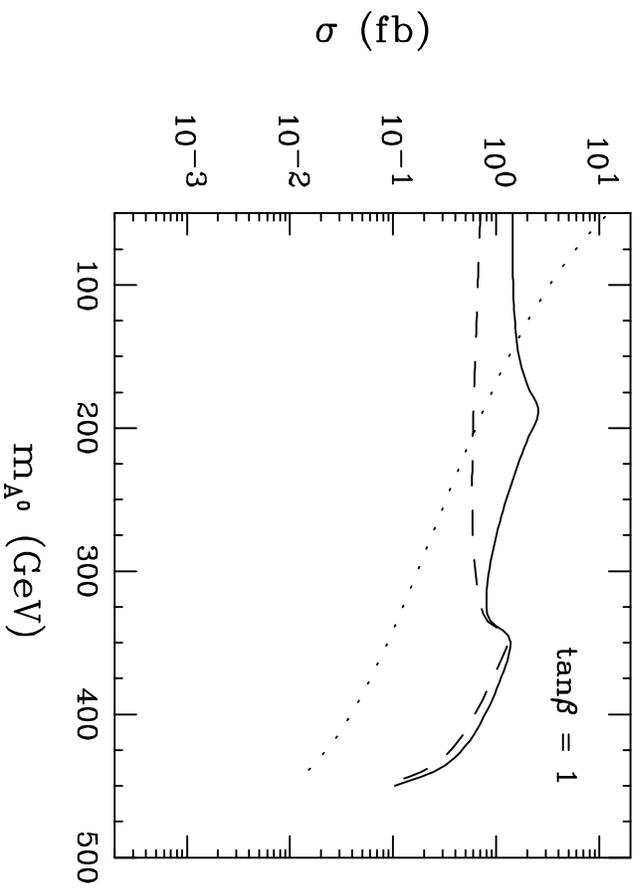
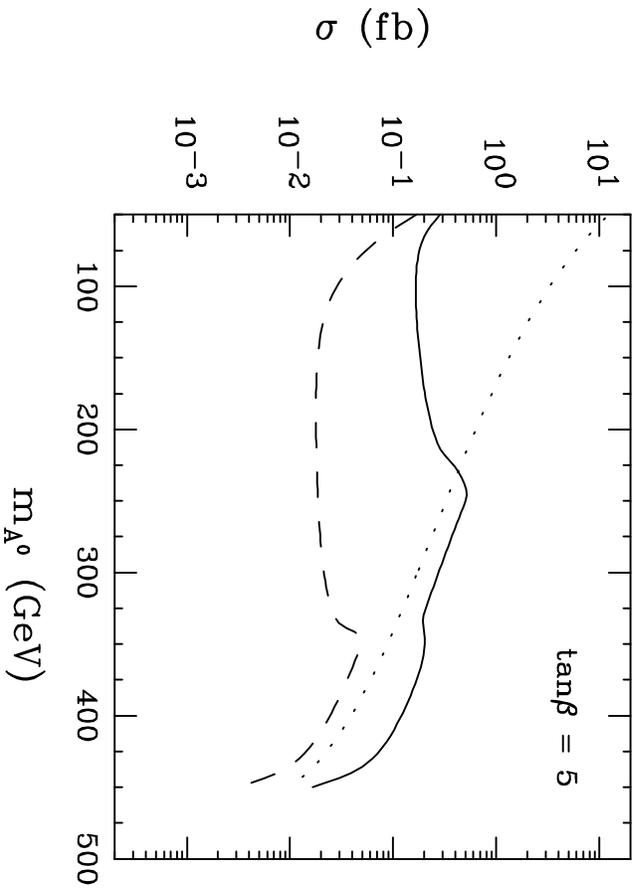
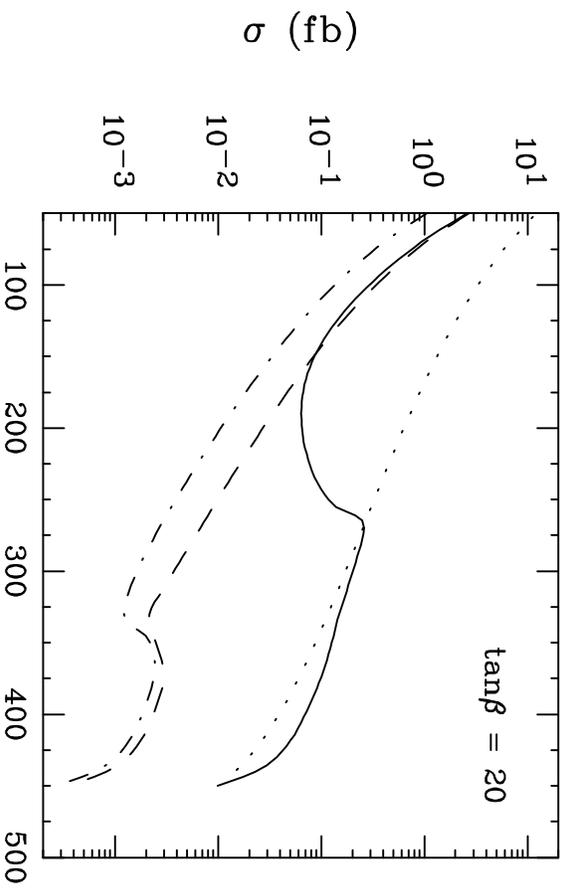
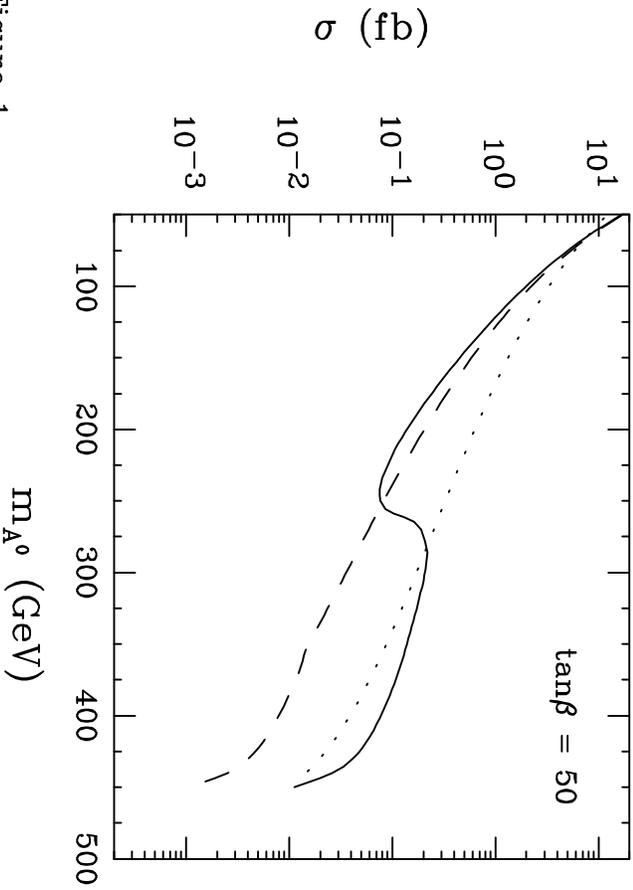
\begin{figure}[h]
\caption{Cross sections for the production of $A^0$ are shown for various values
of $\tan\beta$ and an $e\,\bar{e}$ center of mass energy of 500 GeV when
$M\mu > m_W^2\sin2\beta$. In each case, the solid line corresponds to chargino 
masses $m_1 = 250$ GeV and $m_2$ the largest value consistent with 
the restriction $(m_1 - m_2)\geq m_W\protect\sqrt{2(1 + \sin2\beta)}$. The
dashed line is the standard two Higgs doublet contribution without charginos,
and the dot-dashed line in the $\tan\beta = 20$ panel is the two Higgs doublet
result without the $\tau$ contribution. The dotted lines are the cross section
for the production of a background $b\bar{b}$ with invariant mass $m_{A}$. 
In these graphs, the angular cutoff $\eta$ is taken to be $10^{-5}$.}
\end{figure}

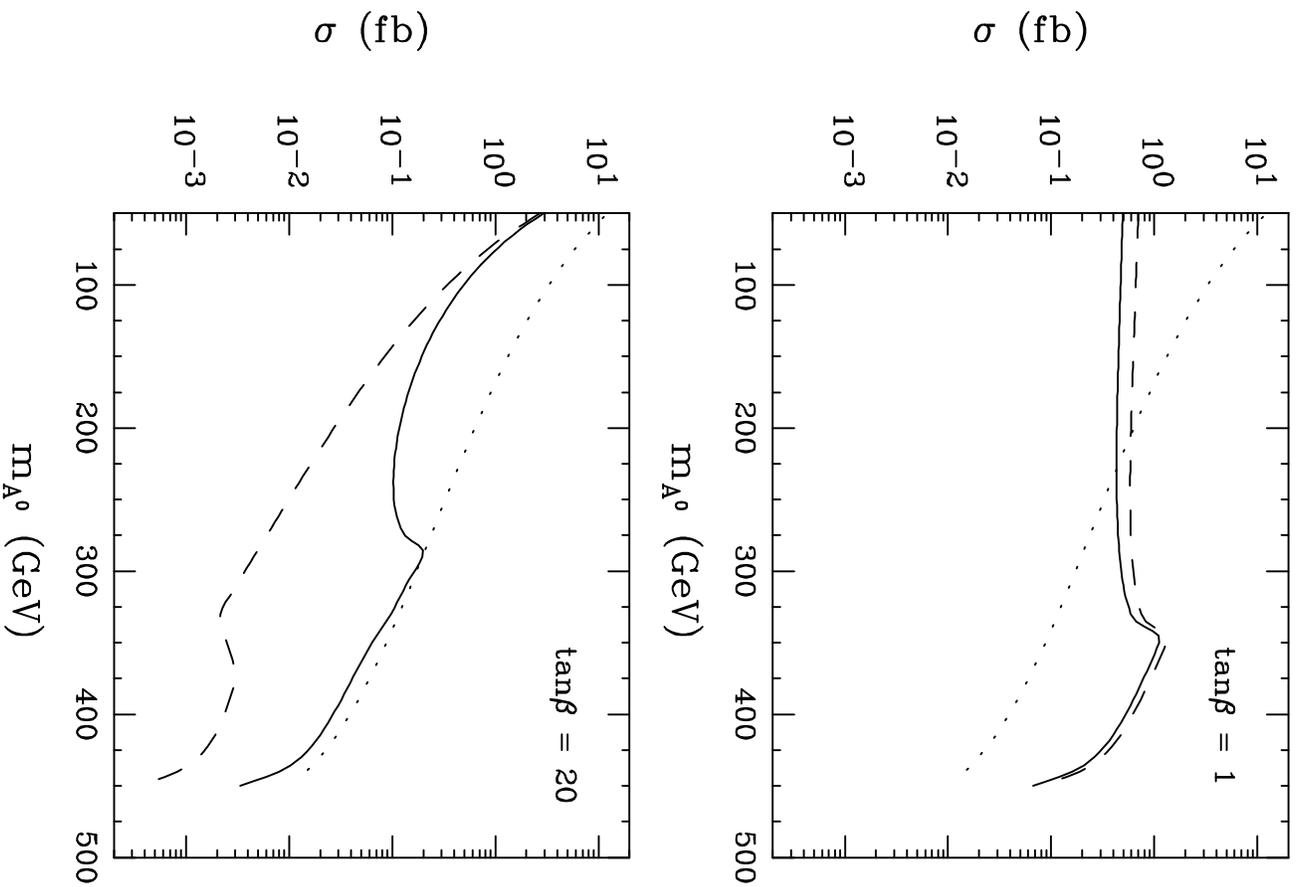
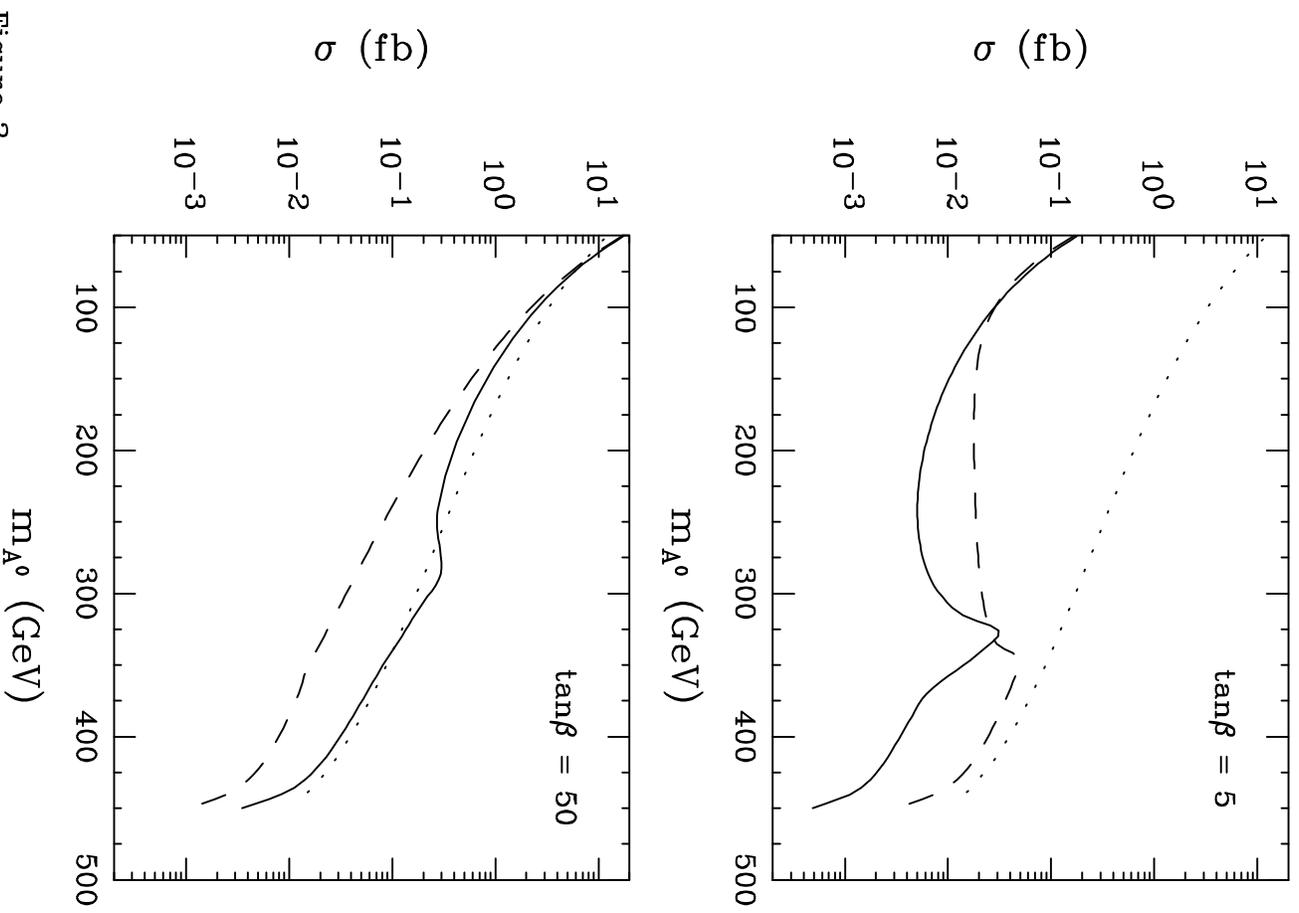
\begin{figure}[h]
\caption{Same as Fig.\,(1) for $M\mu < m_W^2\sin2\beta$. In this case, $m_1$ and
$m_2$ satisfy the condition $(m_1 - m_2)\geq m_W\protect\sqrt{2(1 - \sin2\beta)}
$.}
\end{figure}


\begin{thebibliography}{99}
\bibitem{scalar} The same thing can be said about the production of scalar
Higgs-bosons in the reaction $e\gamma\rightarrow e\,H^0$.
\bibitem{Z} If the $A^0$ mass should turn out to be very close to the $Z$ mass,
there could be a substantial background from $\gamma e\rightarrow Z\,
e\rightarrow b\bar{b}\,e$. We have not included this in our background
calculation.
\end{thebibliography}
\end{document}